\documentclass[oldversion]{aa}

\usepackage{graphicx}

\usepackage{txfonts}

\usepackage{color}

\usepackage{natbib}

\usepackage{color}

\newcommand{\qm}[1]{``#1''}
\newcommand{\fromto}{-}
\def\sss{\scriptscriptstyle}
\def\U{{\sss \!U}}
\def\L{{\sss \!L}}
\def\K{{\sss \!K}}

\def\RP{\mathrm{\scriptstyle{R\!P}}}
\def\nur{\nu_\mathrm{r}}
\def\nuv{\nu_\theta}
\def\nuL{\nu_\L}
\def\nuU{\nu_\U}
\def\nuK{\nu_\K}

\begin{document}

\title{ {Confronting the models of 3:2 quasiperiodic oscillations with the rapid spin of the microquasar GRS~1915+105}}

\author
{Gabriel T\"or\"ok\inst{\nabla}, Andrea Kotrlov\'a\inst{\nabla}, Eva \v{S}r\'amkov\'a\inst{\nabla}, Zden\v{e}k Stuchl\'{\i}k\inst{\nabla}}

\institute{$\nabla$ Institute of Physics, Faculty of Philosophy and Science, Silesian
  University in Opava, Bezru\v{c}ovo n\'{a}m. 13, CZ-74601 Opava, Czech Republic
 }
  
\date{Received / Accepted}

\authorrunning{G. T\"or\"ok et al.}
\titlerunning{ {Models of the 3:2 QPOs vs. the spin of GRS~1915+105}}
 
\date{Received / Accepted}

\abstract
{ {Spectral fitting of the spin $a\equiv cJ/GM^2$ in the microquasar GRS 1915+105 estimate values higher than $a=0.98$. However, there are certain doubts about this (nearly) extremal number. Confirming a high value of $a>0.9$ would have  significant concequences for the theory of high-frequency quasiperiodic oscillations (HF QPOs).  Here we discuss its possible implications assuming several commonly used orbital models of 3:2 HF QPOs. We show that the estimate of $a>0.9$ is almost inconsistent with two hot-spot (relativistic precession and tidal disruption) models and the warped disc resonance model. In contrast, we demonstrate that the epicyclic resonance and discoseismic models assuming the c- and g- modes are favoured. We extend our discussion  to another two microquasars that display the 3:2 HF QPOs. The frequencies of these QPOs scale roughly inversely to the microquasar masses, and the differences in the individual spins, such as $a=0.9$ compared to $a=0.7$, represent a generic problem for most of the discussed geodesic 3:2 QPO models. To explain the observations of all the three microquasars by one unique mechanism, the models would have to accommodate very large non-geodesic corrections.}}

\maketitle

\section{Introduction}
\label{section:introduction}
 

In the past few years an impressive amount of work has been done on estimating the black hole spin using the X-ray continuum fitting method \citep[see, e.g.,][]{mcc-etal:2006,mcc-etal:2008,mid-etal:2006,don-etal:2007,sha-etal:2008,mcc-etal:2010,mcc-etal:2011}. The obtained results indicate that the spins of the individual sources cover almost the whole range spanning from a non-rotating Schwarzschild black hole to near-extreme Kerr black holes \citep[][]{mcc:2010,mcc-etal:2010,mcc-etal:2011}. The latter type of object has been detected in the microquasar GRS 1915+105 for which it was estimated by \cite{mcc-etal:2006} that the dimensionless spin $a\equiv cJ/GM^2$ is higher than $a=0.98$. However, some doubts remain about the (nearly) extremal value reported by McClintock~et~al. In particular, using a similar fitting procedure \cite{mid-etal:2006} derived a value of $a=0.7$. Recent independent spin estimates of GRS 1915+105 based on another, so called, relativistic iron line profile fitting imply the spin either around 0.6 or 0.98 \citep{blu-etal:2009}. There are several concerns about both of the spectral (continuum and iron-line) spin estimate methods. It is not the purpose of our paper to list and analyze these reasons in detail but it is useful to recall some findings of a direct relevance to observations of  low-mass X-ray binaries (LMXBs).

Within continuum method studies, the X-ray spectrum is assumed to be the sum of black body contributions from different disc radii. Within the adopted concept, the expected spectrum is determined by the radial distribution of disc temperature, which follows from the relativistic version of the Shakura-Sunyaev accretion disc model and depends on the black hole spin \citep{sha-sun:1973,nov-tho:1973}. The spin is then inferred by fitting the observed disc spectra to those calculated. In more detail, one of the key theoretical assumptions of this method lies in the expectation that the inner edge of the accretion disc coincides with the innermost stable circular orbit (ISCO). Consequently, the spectral fitting provides an estimate of the ISCO position that depends on the black hole spin \citep[see][for a more complex picture]{sha-etal:2006,mcc-rem:2009,pen-etal:2010}.

The method deals only with a subset of observations where the assumption of the blackbody emission seems valid. Thus, only datasets that show (presumably) weakly Comptonized spectra acquired for the thermal dominant source state have been used in several works \citep[e.g.,][but see also \cite{ste-etal:2009}]{mcc-rem:2003,mcc-rem:2009,sha-etal:2008}. Moreover, only observations where the source luminosity is less than $\sim20\%$ of the Eddington luminosity provide reliable results, since at high luminosities the standard ISCO concept is not valid \citep[see][]{don-dov:2008,abr-etal:2010}. The method is also sensitive to the accuracy of the source distance, mass, and inclination determined from independent methods (usually from the optical measurements).

The iron-line method studies are based on the evidence of a broad, skewed iron-line profile that is also observed in the spectra of several LMXBs. The observed iron line is believed to originate from the reflection of hard X-ray photons in the innermost parts of the accretion disk.  These photons, emitted by an external process (often assumed to be  inverse-Compton scattering in a hot corona), have sufficient energy to remove the K-shell electrons from the iron atoms in the accreted gas and produce the Fe-K emission. The broad profile of the Fe-K line is then explained by  relativistic Doppler effects, light bending, and the gravitational redshift characteristic of the vicinity of the binary central compact object \citep[see][for a review]{mil:2007}. As for spectral continuum fitting, it is assumed that the disc extends down to ISCO. Consequently, properties of the broadened emission line are associated with the ISCO radii when considering the spin as a free parameter. The advantage of this method is that it does not require accurate information about the source distance and mass, and provides estimate of the source inclination itself \citep[e.g.,][]{dav-etal:2006,mcc-rem:2009}. Its weak points are connected namely to important requirements about the strength of the detected signal and to difficulties in subtracting the line from the whole spectra (when adopting detailed modeling of the continuum). Moreover, there are doubts  that the line broadening is caused by general relativistic effects, since these strongly depend not only on the radius of the emission, but also on details of the local emission that are not very known \citep[e.g.,][]{bec-don:2004}.

Both of the aforementioned means of estimating the black hole spin rely on being able to relate the source states to the disc geometry. In the case of GRS 1915+105, it is unclear how this can be achieved \citep[e.g.,][]{mcc-rem:2003,don-etal:2004,oer-etal:2010,li-etal:2010}.
In view of all the uncertainties,  one should keep in mind that some caution is needed in relation to the high value of the black hole spin in GRS~1915+105.\footnote{\cite{mcc-etal:2011} states that $a>0.98$ but the authors also note that some caution is needed namely because of the incomplete study of the source distance (see their paper for details).  In this paper, we show that confirming it would have a significant impact on the theory of the high-frequency quasiperiodic oscillations (HF QPOs) and present the discussion of some concrete implications for a few frequently quoted QPO models.}

\section{Orbital models of quasiperiodic oscillations}
\label{section:models}

The X-ray power density spectra (PDS) of several LMXBs obtained during the past three decades contain peaked features called quasi-periodic oscillations (QPOs). These QPOs are observed in neutron star (NS) as well as black hole (BH) systems and occur with periods of the order of $1\fromto10^{-3}$sec. An outstanding example of these oscillations arises in the high frequency part of the PDS being thus called \qm{high frequency} (HF) QPOs. When more HF QPOs are observed in black hole sources, their frequencies seem to form commensurable pairs with a preferred ratio of 3:2 \citep[][]{abr-klu:2001,mcc-rem:2003}. The black hole HF QPOs are typically weaker and less coherent than most of the NS HF QPOs \citep[e.g.,][]{bar-etal:2005a,bar-etal:2005b,men:2006}. The neutron star HF QPOs appear as two correlated modes at distinct (but, in contrast to BH HF QPOs, variable) frequencies. The range of frequencies spanned by the two NS modes is rather large (ranging from $50$Hz to $1500$Hz, being typically a few hundreds of Hz for each individual source). The two NS modes are in a way analogous to the BH case because they usually exchange their dominance when passing the 3:2 frequency ratio \citep[][]{tor-etal:2008a,tor-etal:2008b,tor-etal:2008c,tor:2009,bou-etal:2010}. Properties of the HF QPOs are reviewed in, e.g., \cite{mcc-rem:2003} or \cite{kli:2006}.

There is strong evidence that the HF QPOs originate very close (less than 100 gravitational radii $r_{\mathrm{g}}$\,$\equiv$\,\mbox{G$M$c$^{-2}$}) to the accreting compact objects, but no commonly accepted QPO theory has so far been developed \cite[e.g.,][]{kli:2006}. 
Moreover, while there are several concrete, well-described similarities and differences between the black hole and neutron star HF QPO phenomenology (including those briefly mentioned above), it has not yet been resolved whether the generic mechanism could be the same for both classes of the sources.  There are several proposed models that cannot be applied to both the classes, e.g., due to the requirement of a solid NS surface \citep{lam-etal:1985, alp-sha:1985} or a high black hole spin \citep{stu-etal:2007a,stu-etal:2007b,sla-stu:2008}. Nevertheless, a larger variety of models has been designed under the assumption of a common mechanism \citep[see, e.g.,][for a basic overview]{kli:2006}. Most of the hypotheses assume a relation between the QPO frequencies and the frequencies related to motion of accreted matter orbiting in the vicinity of a compact object (hereafter \qm{orbital} models). These models typically deal with either \qm{hot-spot} or \qm{disc-oscillation} QPO interpretation.  {We next consider an arbitrary choice of several commonly quoted models. For the sake of comprehensibility, we first give a short summary of the examined models and recall some of their main features and related references.}

\subsection{ {Kinematic models}}

The kinematics of the orbital motion allows the consideration of the variability that arises from the motion of \qm{hot-spots} orbiting inside the accretion disc. We consider two models of this kind. The \qm{relativistic precession} model (hereafter RP model) was proposed in a series of papers by \cite{ste-vie:1998a,ste-vie:1999,ste-vie:2002,mor-ste:1999}. The model illustrates that the kHz QPOs represent modes of the relativistic epicyclic motion of blobs at various radii $r$ in the inner parts of the accretion disc. It is often recalled and known for roughly matching the correlation between the HF QPOs observed in the NS sources \citep[e.g.,][]{bel-etal:2007a,tor-etal:2010,lin-etal:2010}. Within the model, the twin-peak QPO frequency correlation arises because of the periastron precession of the relativistic orbits. Owing to Lense-Thirring relativistic precession, the model also predicts another frequency correlation extending to higher timescales. The kHz QPO frequencies are indeed correlated to the low-frequency QPO features observed around $\sim\!$\,1--50Hz \citep[e.g.,][this correlation is however not being the focus of our present work]{ste-vie:1998b}. On the other hand, there are apparent difficulties with the detailed modeling of the relation between the hot-spot motion and the observed modulation \citep[e.g.,][]{lam-mar:2000}. 

\cite{cad-etal:2008}, \cite{kos-etal:2009}, and \cite{ger-etal:2009} introduced a similar concept in which the QPOs are generated by a \qm{tidal disruption} (TD) of large accreting inhomogeneities. Their hydrodynamic simulations signify that blobs orbiting the central compact object can be stretched by tidal forces forming  \qm{ring-section} features within the model  expected to be responsible for the observed modulation. The model has been proposed for both supermassive and stellar mass black holes. We note that, at least in some cases, the power density spectra simulated using  the model closely reproduce those that are observed. However, the concept of "rocks" approaching the radii of the order of ISCO-radius in LMXBs is questionable. The upper limit to the radii, where the tidal forces begin to disrupt the object of  density $\rho$ orbiting a primary source of mass $M$, is roughly described by the Roche-limit $r_{\mathrm{TD}}\sim (M/\rho)^{1/3}$. The related characteristic orbital frequency  $\nu_{\mathrm{TD}}\sim(\mathrm{G}M/r_{\mathrm{TD}}^3)^{1/2}$ can be expressed as $\nu_{\mathrm{TD}}=(\mathrm{G}\rho)^{1/2}$. For the rocks it is clearly given by $\nu_{\mathrm{TD}}\sim10^{-3}$Hz, which is far lower than the observed ($\sim100$Hz) frequencies.\footnote{We thank the anonymous referee for emphasizing this issue.} 

Identification of the lower and upper kHz QPO frequencies with the frequencies of the orbital motion is recalled in the top part of Table~\ref{table:models} for both hot spot models. To match the observed 3:2 ratio, the QPOs in the RP model must be generated at the radii where $\nuK/\nur=3/1$ \mbox{($r=6.75{\mathrm{G}M\mathrm{c}^{-2}}$ for $a=0$)}, i.e., very close to ISCO \mbox{($r=6{\mathrm{G}M\mathrm{c}^{-2}}$ for $a=0$)} where the radial epicyclic frequency vanishes. For the TD model, this location is shifted outwards close to the radius where the radial epicyclic frequency exhibits a maximum value ($\nuK/\nur=2/1$; $r=8{\mathrm{G}M\mathrm{c}^{-2}}$ for $a=0$).

\subsection{{Resonant models}}

Within the (present) RP model, there is no generic explanation of the observed 3:2 frequency ratio $R=3/2$ and there is a clear need to explore the issue of the preference of certain orbits. The concept of the TD model is less problematic because for any $a$ and $r$ it implies that $R\in(1,\,2)$ and the predicted effects occur around the location of the maximal allowed $\nur$, where $\nuK/\nur\sim2/1$. Nevertheless, the frequency ratio is also not reliably constrained by this model. {In contrast, the commensurability of the frequencies is crucial for the models when assuming the warped disc (WD) oscillations suggested by \cite{kat:2001,kat:2004a,kat:2004b,kat:2005,kat:2008}.} \cite{kat:2008} reviewed several possible resonances in deformed disc \citep[see also works of][]{fer-ogi:2008,fer-ogi:2009}. The main weak point of the concrete WD model assumed here \citep[][see Table\ref{table:models}]{kat:2004a,kat:2004b} is that it considers a somewhat exotic disc geometry that causes a doubling of the observed lower QPO frequency.

The frequency commensurability is also crucial for the non-linear resonance models discussed by Abramowicz, Klu{\'z}niak and collaborators \citep[][and others; see also \citeauthor{ali-gal:1981}, 1981 and \citeauthor{ali:2006}, 2007]{klu-abr:2001,abr-klu:2001,abr-etal:2003a,abr-etal:2003b,bur-etal:2004,reb:2004,hor:2008,stu-etal:2008a,stu-etal:2008b,hor-etal:2009}. A particular resonance of this kind that is often discussed is the 3:2 internal epicyclic or Keplerian resonance \citep[Ep, Kp, see][for details]{tor-etal:2005}. The explicit formulae of the frequency relations corresponding to these three resonant disc oscillation models are listed in the bottom part of Table \ref{table:models}. We note that while for the WD resonance model the QPOs are located at the same radii as for the kinematic TD model, i.e., around the location of the maximum of $\nur$, the two 3:2 resonances occur above this maximum, at a larger distance from the central object. We also consider here another two QPO (resonance) models that we denote as RP1 model \citep{bur:2005} and  RP2 model \citep{tor-etal:2010}. Both of them assume different combinations of non-axisymmetric disc-oscillation modes. They are of particular interest because they involve oscillation modes whose frequencies for slow rotation almost coincide with the frequencies predicted by the RP model. These two resonances should occur much closer to ISCO than the resonance expected in the Ep model.

 {The physical interpretation of the RP1 model is unclear, since the oscillation modes assumed within the model are unlikely to be able to enter the resonance \citep[see,][for details]{hor:2008}. The resonant coupling between the pairs of the oscillation modes assumed in the Kp, EP, and RP2 models is in principle allowed, but detailed  physical mechanisms providing this coupling and mode excitations have not yet been fully developed \citep[see, e.g.,][and references therein]{sra-etal:2007,hor:2008,reb:2008,klu:2008}. 
Substantial effort clearly yet needs to be invested in extending both the related analytic and numerical work and the link between them.}

\subsection{{Models assuming fundamental discoseismic modes}}

 {We have so far only considered models in which it is assumed that both of the observed 3:2 frequencies are produced by the same mechanism and excited at a certain (common) preferred radius. A qualitatively different consideration relating each of the two frequencies to a different radial region arises when general combinations of the fundamental discoseismic modes are assumed.} The HF QPOs have been proposed to correspond to the following three distinct modes: the so-called g-modes (inertial-gravity waves that occur at the radius where the radial epicyclic frequency reaches its maximum value), c-modes (corrugation vertically incompressible waves near the inner edge of the disk), and p-modes (inertial-pressure oscillations that occur near the edge of the disc). This concept was elaborated in particular by   \cite{kat-fuk:1980}, \cite{oka-etal:1987}, \cite{now-wag:1992}, \cite{wag:1999}, \cite{sil-etal:2001}, \cite{wag-etal:2001}, \cite{ort-etal:2002}, \cite{sil-wag:2008}, and \cite{wag:2008}. Disc oscillation modes that have been proposed to explain the QPOs have been observed in hydrodynamical (HD) simulations of the accretion processes \citep[e.g.,][]{zan-etal:2005,rey-col:2009} but there is growing evidence from magnetohydrodynamic (MHD) simulations that these modes are typicaly dumped by various instabilities produced in the presence of the magnetic field \citep[see, e.g.,][]{tsa-lai:2009,fu-lai:2009,fu-lai:2011}.  {At the same time, the MHD simulations do not convincingly reproduce the 3:2 QPOs that undoubtedly appear in the X-ray fluxes of the LMXBs. It is therefore also unclear whether these simulations accomodate all the crucial ingredients or not \citep[see, e.g.,][who found in their MHD simulations some low frequency QPOs only in the presence of the radiative cooling]{mac-mat:2007,mac-mat:2008}.}

\subsection{Confronting the QPO models with the rapid spin of black hole in the microquasar GRS~1915+105}

 {In general, the existing QPO models represent rather unfinished concepts that have specific advantages and difficulties. The oscillations predicted by these models are usually not seen in the present MHD simulations. Moreover, none of these models yet really match the full QPO phenomenology observed in the LMXBs including the dependence of the QPO visibility on the source spectral states, QPO amplitudes, and coherence times. In this situation, we expect that the implications of the eventually confirmed high spin of GRS~1915+105 could  help us improve the individual QPO models or even discriminate between them. In the next section, we calculate the spin implied by the above-mentioned QPO models.}

\begin{figure*}[t!]
\begin{center}
\includegraphics[width=1\hsize]{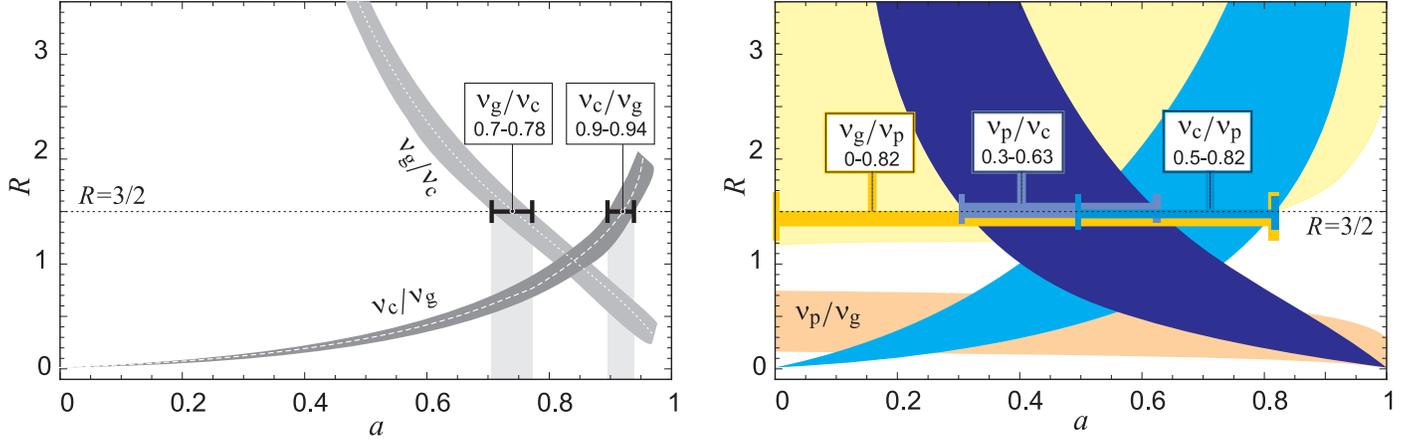}
\end{center}
\caption{Ratio of the frequencies of the fundamental discoseismic modes \citep[based on][]{wag-etal:2001}. Left: Ratio of $\nu_\mathrm{c}$ to $\nu_\mathrm{g}$. The spread of two functions corresponds to the uncertainty in the speed of sound. The horizontal dotted line denotes the 3:2 frequency ratio. Values in the boxes evaluate the spin ranges required by this ratio. Right: The same consideration, but including the p-modes.}
\label{figure:disco}
\end{figure*}

\begin{table*}
\caption{Frequency relations corresponding to individual QPO models and the spins implied by the 3:2 QPOs in GRS 1915+105 and the mass-range $10\fromto18M_{\sun}$. The middle column indicates the ratio of the epicyclic frequencies determining the radii corresponding to the observed 3:2 ratio. The indicated ranges of spin also represent total spin ranges for the whole group of the three microquasars (see Figure~\ref{figure:microL}).}
\label{table:models}
\begin{center}
\renewcommand{\arraystretch}{1.35}
\begin{tabular}{lllcr}
    \hline
  \hline
     \textbf{Model} & \multicolumn{2}{c}{\textbf{Relations}} & $\mathbf{\nu_\mathrm{K}/\nu_\mathrm{r}}$~\textbf{or~}$^*\mathbf{\nu_{\theta}/\nu_\mathrm{r}}$&~~a$\sim$\\
    \hline
$\mathbf{RP}\phantom{1}$ & $\nu_{\mathrm{L}} =\nu_{\mathrm{K}}-\nu_{\mathrm{r}}$ & $\nu_{\mathrm{U}} =\nu_{\mathrm{K}}$ & $3/1^*$ & $<$0.55\\
$\mathbf{TD}\phantom{1}$ & $\nu_{\mathrm{L}}=\nu_{\mathrm{K}}$ & $\nu_{\mathrm{U}}=\nu_{\mathrm{K}}+\nu_{\mathrm{r}}$ & 2/1$\phantom{^*}$&$<$0.45\\
\hline
$\mathbf{WD\phantom{1}}$ & $\nu_{\mathrm{L}}=2\left(\nu_{\mathrm{K}}-\nu_{\mathrm{r}}\right)$ & $\nu_{\mathrm{U}}=2\nu_{\mathrm{K}}-\nu_{\mathrm{r}}$ & 2/1$\phantom{^*}$& $<$0.45\\
$\mathbf{Ep\phantom{1}}$ & $\nu_{\mathrm{L}}=\nu_{\mathrm{r}}$ & $\nu_{\mathrm{U}}=\nu_{\theta}$ & $3/2{^*}$& 0.65 -- 1\\
$\mathbf{Kp\phantom{1}}$ & $\nu_{\mathrm{L}}=\nu_{\mathrm{r}}$ & $\nu_{\mathrm{U}}=\nu_{\mathrm{K}}$ & $3/2^*$ & 0.70 -- 1\\
$\mathbf{RP1\phantom{1}}$ &  $\nu_{\mathrm{L}}=\nu_{\mathrm{K}}-\nu_{\mathrm{r}}$ &  $\nu_{\mathrm{U}}= \nu_{\theta}$ &--$\phantom{^*}$& $<$0.80\\
$\mathbf{RP2\phantom{1}}$ &  $\nu_{\mathrm{L}}=\nu_{\mathrm{K}}-\nu_{\mathrm{r}}$ &  $\nu_{\mathrm{U}}= 2\nu_{\mathrm{K}}-\nu_{\theta}$ & --$\phantom{^*}$& $<$0.45\\
\hline
\end{tabular}
\end{center}
\end{table*}

\section{The spin implied by individual models}

There is a straightforward connection between the hot-spot models and the characteristic orbital frequencies (i.e., the three frequencies of the perturbed circular orbital motion: the azimuthal \qm{Keplerian} frequency, the radial epicyclic frequency, and the vertical epicyclic frequency). The frequencies of numerous  disc-oscillation modes are also constrained by the three characteristic orbital frequencies. Assuming a Kerr geometry, the orbital frequencies for a given radius depend only on mass and spin of the black hole.\footnote{We consider here only the Kerr spacetime as a standard description for rotating BHs, although alternatives have been discussed in a similar context \citep[see][]{kot-etal:2008,stu-kot:2009}.}  {It is therefore possible to infer the black hole spin or mass from the observed 3:2 frequencies and (at least some) concrete orbital models. This procedure has been previously performed, e.g., for a wide selection of the resonance QPO models \citep{abr-klu:2001,tor-etal:2005,tor:2005}. We recall that the complete set of formulae required to evaluate the Keplerian and epicyclic orbital frequencies in Kerr geometry was first derived by \cite{ali-gal:1981}. These frequencies were extensively discussed in several later studies. A detailed analysis can be found in \cite{tor-stu:2005}, where the definition formulae are given in the usual form and the frequencies are studied across the full range of positive $a$. We note that here we do not consider $a>1$ since we focus our attention fully on models originally designed for BHs and not on their possible extensions to naked singularities.}

The 3:2 QPO frequencies in GRS~1915+105 are well known to be given by \citep[][]{mcc-rem:2003}
\begin{equation}
\label{equation:3:2}
    \nu_{\mathrm{U}}\!=\!168\pm3\mathrm{Hz~and~}\nu_{\mathrm{L}}\!=\!113\pm5\mathrm{Hz}.
\end{equation}
Assuming Eq. (\ref{equation:3:2}) and the aforementioned formulae for the orbital frequencies, we calculate the implied mass-spin functions for the models associating the 3:2 QPOs with a common radii by means of the definition relations given in Table~\ref{table:models}. Following \cite{abr-klu:2001} and \cite{tor-etal:2005} and taking into account the estimated range of the mass of GRS~1915+105 \citep[e.g.,][]{mcc-rem:2003}
\begin{equation}
\label{equation:mass}
    10\,\mathrm{M}_{\odot} \le M \le 18\,\mathrm{M}_{\odot},
\end{equation}
we infer the expected ranges of the spin.
We found that the RP, TD, WD and RP2 models imply that the spin is rather too low, i.e. $a<0.6$. Somewhat more satisfactory are the predictions obtained from the Ep ($0.65\le a \le 1$), Kep ($0.7\le a \le 1$), and RP1 ($a \le 0.8$) models. For each model, the detailed results are presented in the last column of Table~\ref{table:models}. 
As recalled above, for the discoseismic modes the individual observed QPOs correspond to different modes located at their own radii. The frequencies of these modes depend on the black hole spin and the speed of sound in the accreted gas, and scale roughly as $1/M$, whereas their dependence on the other parameters of the  accreting system is supposed to be very weak \citep[][]{wag-etal:2001}. The frequencies of the c- and g- modes ($\nu_\mathrm{c}$ and $\nu_\mathrm{g}$) depend only moderately on the speed of sound. Thus, their ratio is mainly a function of the black hole spin, which is depicted in the left panel of Figure~\ref{figure:disco}.  The right panel of Figure~\ref{figure:disco} shows the relations of the ratios of the frequencies $\nu_\mathrm{c}$ or $\nu_\mathrm{g}$ to the p- mode frequency $\nu_\mathrm{p}$, which are  strongly dependent on the speed of sound. Inspecting both panels of  Figure~\ref{figure:disco}, we can see that for the combination of modes relating $\nuU=\nu_\mathrm{p}$ and $\nuL=\nu_\mathrm{g}$ the frequency ratio $R$ is never higher than unity. For the other five possible combinations of modes, there is in each case some range of $a$ that corresponds to the 3:2 frequency ratio. This range is located above $a=0.9$ only for the combination relating $\nuU=\nu_\mathrm{c}$ and $\nuL=\nu_\mathrm{g}$. In more detail, this combination requires that $0.9\le a\le0.94$.\footnote{For higher values of $a$, the model predictions seem to deviate from the 3:2 ratio since the c-mode frequencies, when $a$ increases, grow faster than the frequencies of the g-mode. However, the published works have not fully investigated the behaviour for very high spin values $a>0.95$ so far because of numerical difficulties (Wagoner 2010, from a private communication).} The mass needed to match $\nu_{\mathrm{U}}=168\pm3\mathrm{Hz}$ is then $13\mathrm{M}_{\odot} \le M \le 19\mathrm{M}_{\odot}$ with good overlap $13\mathrm{M}_{\odot} \le M \le 18\mathrm{M}_{\odot}$ with the expected mass given in Eq. (\ref{equation:mass}).

\section{Discussion and conclusions}
\label{section:conclusions}

\begin{figure}[t]
\begin{center}
\includegraphics[width=1\hsize]{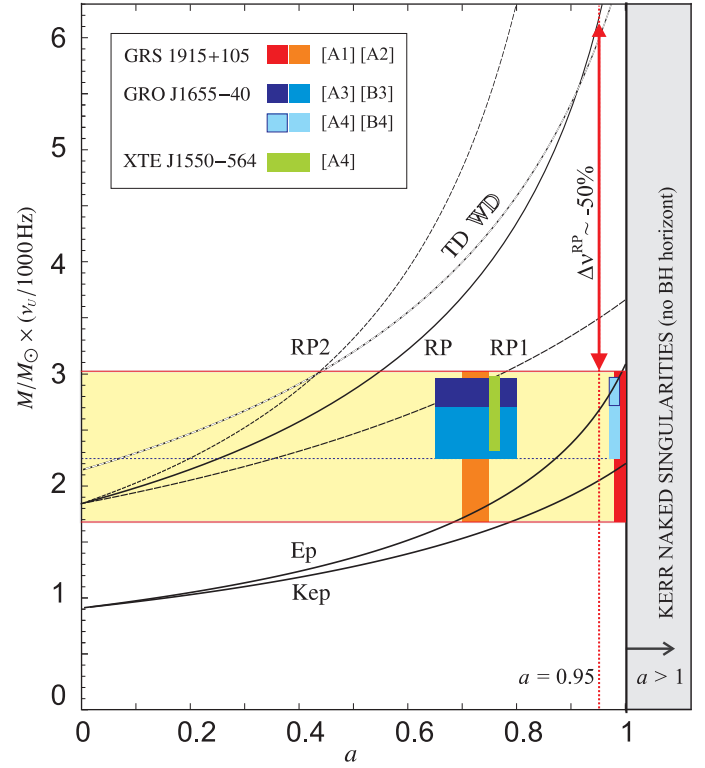}
\end{center}
\caption{Curves $\nuU\times M(a)$ implied by the individual geodesic models. The light yellow rectangle indicates the observationaly determined interval of $\nuU\times M$ for GRS~1915+105. The red dotted vertical line denotes $a=0.95$. The red vertical arrow indicates the correction needed to match the upper limit to $\nuU\times M$ with the RP model for this spin.  The colour boxes are drawn for the mass and spectral spin estimates given by different authors for GRS 1915+105, GRO J1655-40 and XTE 1550-564 (see the main text for references). The dotted blue line indicates the lower observational limit to $M\times \nuU$ that is roughly common to GRO J1655-405 and XTE 1550-564.}
\label{figure:microL}
\end{figure}

It follows from our previous discussion that the hypothesis of the internal (epicyclic) resonance\footnote{\citet{tor-etal:2005} assumed this and several other resonances for the spin estimates of the Galactic microquasars. From the other considered resonances,  the \qm{5:1} resonance can also match the high spin. This resonance involves  relatively high resonant coefficients and occurs very close to ISCO.} 
and the discoseismic model which relates the upper and lower 3:2 QPO to the c- and g- modes are favoured in the case of GRS~1915+105 provided that $a>0.9$. The TD, WD, RP, and RP2 models are then disfavoured. This statement was inferred assuming that $\nuK$, $\nur$, and $\nuv$ are the exact geodesic frequencies. A similar analysis including the influence of non-geodesic effects would require very detailed study. Here we only roughly estimate the possible relevance of the non-geodesic effects.  {Figure~\ref{figure:microL} shows the $\nuU\times M(a)$ curves implied by the individual geodesic models. These curves (as well as the curves drawn in the consequent Figure~\ref{figure:microR}) have non-trivial ambiguous extensions for the range of $a>1$ representing the naked-singularity region. We do not consider them here, since we focus our attention on black holes (we expect to discuss the subject of $a>1$ in a different work).} The observationally determined interval of $\nuU\times M$ for GRS~1915+105 is indicated by the light yellow rectangle. We define the relative non-geodesic correction to be
\begin{equation}
\Delta\nu\equiv{(\nu_{\mathrm{observed}}-\nu_{\mathrm{predicted}})}/{\nu_{\mathrm{predicted}}},
\end{equation}
which is needed to match the observations of GRS~1915+105  with a given model for a certain spin. From Figure~\ref{figure:microL}, we can easily estimate by eye that the non-geodesic correction $\Delta\nu^{{\RP}}$ required for the frequency predicted by the RP model for $a=0.95$ is about $-50\%$. From Figure~\ref{figure:microR}, we can find that for $a\in(0.9,1)$, the quantity $\Delta\nu^{{\RP}}$  changes from $-40\%$ to $-60\%$. The same is roughly true for the TD and WD models, while for the RP2 model the required correction is even higher. Thus, our result is justified, except when considering of very large non-geodesic corrections.

\subsection{Relation to other sources}

\citet{tor-etal:2005} pointed out that since the 3:2 QPO frequencies in microquasars scale roughly  as \mbox{$\nuU\!\doteq\!2.8(M/M_{\odot})^{-1}$\,kHz} \citep{mcc-rem:2003}, their spins implied by the epicyclic resonance model should not vary much among them.  {In Figure~\ref{figure:microL}, we include several examples of spin estimates obtained by different groups and methods for GRS~1915+105 and the other two microquasars, GRO J1655-40 and XTE 1550-564. The references related to these estimates are as follows \citep[the assumed ranges of $\nuU$ being taken from][]{mcc-rem:2003}.} For the mass,  \cite{gree-etal:2001}, \cite{grei-etal:2001}, \cite{oro-etal:2002}, \cite{bee-pod:2002}, and \cite{mcc-rem:2003} provide commonly accepted mass estimates. These are in Figure~\ref{figure:microL} denoted by letter A. \citet{bee-pod:2002} present an alternative prediction (in Figure~\ref{figure:microL} denoted by letter B) that moves the lower boundary of estimated mass of GRO J1655-40 from $6.0$ to $5.1$ $M_{\odot}$.
For the spin, there were the following studies: 
[1] \cite{mcc-etal:2006}; 
[2] \cite{mid-etal:2006}; 
[3] \cite{mcc-etal:2008}; 
[4] \cite{mil-etal:2009}. 

The above quoted spin estimates assume either the spectral continuum or the iron line method. It is apparent from Figure~\ref{figure:microL} that both different authors and different methods suggest (somewhat) different values of spin (and for one microquasar even different values of mass). The epicyclic resonance model favoured (along with the discoseismic model) in the GRS~1915+105 seems to match at least some of these estimates, but for the parameters of the XTE J1550-564 ($a\!\sim\!0.7$,~$M\!\sim\!5-7M_{\sun}$) assumed in the figure it fails. If different spins ($a>0.9$ in GRS~1915+105 and $a\sim0.7$ in GRO J1655-40 and XTE J1550-564)  were confirmed, the difficulty of matching the all observed 3:2 frequencies would clearly be rather generic for most of the orbital QPO models. 

\begin{figure}[t]
\begin{center}
\includegraphics[width=1\hsize]{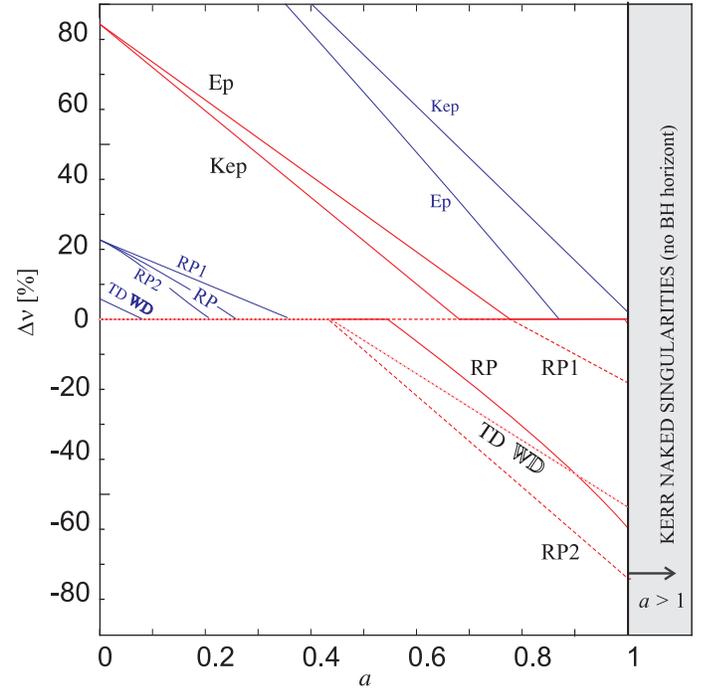}
\end{center}
\caption{Non-geodesic corrections required for a given model, spin and source.  The red curves indicate the minimal corrections required in the case of GRS~1915+105. Parts of these curves with a negative sign also roughly indicate the corrections required for the other two microquasars. The blue curves indicate positive corrections for these microquasars determined by the lower limit to their $M\times \nuU$ indicated by the blue dotted horizontal line in Figure~\ref{figure:microL}.}
\label{figure:microR}
\end{figure}

The observationally determined ranges of $\nuU\times M$ for GRO J1655-40 and XTE J1550-564 nearly coincide, the upper limits being roughly equal to the upper limit for GRS 1915+105 (see Figure~\ref{figure:microL}). Taking advantage of this setup, we plot a simple rough scheme of the non-geodesic corrections required for a given model, spin, and source in Figure~\ref{figure:microR}. The red curves in the figure indicate the minimal corrections required for GRS~1915+105. Parts of these curves with a negative sign also roughly indicate the corrections required for the other two microquasars. Positive corrections for these microquasars are then included as the blue curves.  {If the values $a>0.9$ for GRS~1915+105 and $a\sim0.7$ for GRO J1655-40 or XTE J1550-564 were confirmed, the requested corrections would appear to be rather high for most of the denoted models assuming a unified 3:2 QPO mechanism. Only the RP1 model can survive with corrections of $|\Delta\nu|$ up to $\sim20\%$ (but as recalled in Section~\ref{section:models}, the present physical interpretation of this model is questionable).}

Because of the observational $1/M$ scaling, the above difficulty also arises for the discoseismic models (which are not considered in Figures~\ref{figure:microL}~and~\ref{figure:microR}). For these, we present in Table~\ref{table:disco} the mass ranges implied by all combinations of the fundamental modes. These appear to overlap well with those observationally determined only for the model relating the upper and lower 3:2 QPO to the c- and g- mode provided that $a\in(0.90-0.94)$. For the other combinations related to different spins, the mass ranges differ from those in the observation. Clearly, there is the need for a substantial correction also for a unified 3:2 QPO model assuming fundamental discoseismic modes provided that microquasars had different spins of $a>0.9$ compared to $a\sim0.7$.

\section*{Acknowledgments}
We thank M. Abramowicz, W. Klu{\'z}niak and R. Wagoner for their comments and the anonymous referee for objections and suggestions that helped to greatly improve the paper. We thank to J. McClintock and Hor\'ak, J. for discussion. The authors would also like to acknowledge the support of the Czech grants MSM~478130590, GA\v{C}R 202/09/0772 and LC~06014, and the internal grants of SU Opava, FPF SGS/1,2/2010. Part of the work reported here was carried out during the stay of GT and ZS at the University of Gothenburg which was co-supported by The Swedish Research Council grant (VR) to M. Abramowicz.

\renewcommand{\arraystretch}{1.35}
\definecolor{slightlyGray}{gray}{0.85}
\newcommand {\graybox} [1] {\raisebox{0pt}[0pt]{\fcolorbox{black}{slightlyGray}{#1}}}
\newcommand {\grayarrea} [1] {\colorbox{slightlyGray}{#1}}   
\begin{table*}
\caption{Ranges of $a$ and $M$ implied by the discoseismic models of the 3:2 QPOs. The values in the brackets indicate the referential interval of $M$ for each microquasar. The shadows emphasize the ranges of $M$ having some overlap with the referential values.}
\label{table:disco}
\begin{center}
\begin{tabular}{|c|c|c|c|c|}
  \hline
                &&\textbf{GRS 1915+105} & \textbf{XTE J1550-564} & \textbf{GRO J1655-40}  \\
Frequencies   & ${a}$&       ${M/M_{\sun}~[10.0-18.0]}$ & ${M/M_{\sun}~[8.4-10.8]}$ & ${M/M_{\sun}~[5.1-6.6]}$ \\
  \hline
    \hline
$\nuU=\nu_{\mathrm{g}},~\nuL=\nu_{\mathrm{c}}$ &0.70-0.78         & $6.4 - 9.0$    & $3.9 - 5.5$   &  $2.4 - 3.3$ \\
$\nuU=\nu_{\mathrm{c}},~\nuL=\nu_{\mathrm{g}}$ &\textbf{0.90-0.94}& $\grayarrea{12.8 - 19.1}$  & $\grayarrea{7.8 - 11.6}$  &  $\grayarrea{4.8 - 7.1}$ \\
\hline
$\nuU=\nu_{\mathrm{p}},~\nuL=\nu_{\mathrm{c}}$ &0.30-0.63         & $1.1 - 5.0$    & $0.7 - 3.0$   &  $0.4 - 1.8$ \\
$\nuU=\nu_{\mathrm{c}},~\nuL=\nu_{\mathrm{p}}$ &0.50-0.82         & $1.7 - 8.6$    & $1.0 - 5.2$   &  $0.6 - 3.2$ \\
\hline
$\nuU=\nu_{\mathrm{g}},~\nuL=\nu_{\mathrm{p}}$ &0.00-0.82         & $3.9 - 9.7$    & $2.4 - 5.8$   &  $1.5 - 3.6$ \\
$\nuU=\nu_{\mathrm{p}},~\nuL=\nu_{\mathrm{g}}$ & ----   & ----   &  ---- & ---- \\
\hline
\end{tabular}
\end{center}
\end{table*}



\end{document}